\documentclass[prl,twocolumn ,superscriptaddress]{revtex4}
\usepackage{amssymb} 
\usepackage{graphicx}
\usepackage{epsfig}  
\usepackage{subfigure} 
\usepackage{psfrag}

\begin{document} 

 
\title{Enumeration of RNA structures by Matrix Models} 
 
\author{Graziano Vernizzi} 
\affiliation{Service de Physique Th\'eorique, CEA Saclay, 91191 
Gif-sur-Yvette Cedex, France}  
 
\author{Henri Orland} 
\affiliation{Service de Physique Th\'eorique, CEA Saclay, 91191 
Gif-sur-Yvette Cedex, France}

\author{A. Zee} 
\affiliation{Department of Physics, University of California, Santa 
Barbara, CA 93106, USA} 
\affiliation{Institute for Theoretical Physics, University of 
California, Santa Barbara, CA 93106, USA} 
 
\begin{abstract} 
We enumerate the number of RNA contact structures
according to their genus, i.e.~the topological character of their
pseudoknots. By using a recently proposed matrix model formulation for
the RNA folding problem, we obtain exact results for the
simple case of an RNA molecule with an infinitely flexible backbone,
in which any arbitrary pair of bases is allowed.  We analyze the
distribution of the genus of pseudoknots as a function of the total
number of nucleotides along the phosphate-sugar backbone.
\end{abstract} 
\maketitle 
 
The prediction of foldings of single-stranded nucleic acids (like RNA
molecules) is still a major open problem of molecular biology \cite{tinoco}.
Several methods are available for the prediction and description of
the folding process in various conditions. Most of them are
statistical models (both at equilibrium and out-of equilibrium) that
have roots in combinatorial problems. Although
these models are much simpler than the energy based ones (and thus
cannot provide thermodynamical predictions), 
they often provide exact analytical solutions that give important
insights on the phase-space structure and the entropy. For those
reasons the combinatorics of contact structures of biopolymers has
received great attention over the past thirty years~\cite{waterman}. 
In the case of
RNA-folding, a lot of attention has been paid to the combinatorics
of contact structures that are planar (see e.g.~\cite{combinatorics1}
or \cite{combinatorics2} and references therein), but very little is 
known about non-planar structures (i.e.~structures with pseudoknots).  
In this Letter we
explore a very schematic model for RNA folding which 
allows for the exact enumeration of all contact structures with fixed genus.
This model, which is based on a simpler one that was proposed earlier in
\cite{oz,POZ,PTOZ,VOZ}, may be relevant for studying the
behaviour of non-planar contributions.  The partition function is that
of a chain of $L$ nucleotides in three dimensions:
\begin{equation} 
\mathcal{Z} = \int \prod_{k=1}^L d^3 \mathbf{r}_k  
           f(\{ \mathbf{r} \} ) Z_{L}(\{ \mathbf{r} \} ) \, , 
\end{equation}  
where $\mathbf{r}_k$ is the position vector in three dimensions of the
$k$-th base, and $f(\{ \mathbf{r} \} )$ is a function which takes into
account the geometry, the stiffness and the sterical constraints of
the chain. The folding of the chain is caused by the hydrogen bonds
that the bases can form. Since the hydrogen bonds saturate,
a base can interact with only one other base at a time.  
The contribution from such interactions to the
partition function is described by
$Z_{L}(\{ \mathbf{r} \} )$:
\[
  Z_{L}(\{ \mathbf{r} \} ) 
 =  1+  \sum_{i<j} V_{ij}(\mathbf{r}_{ij}) +
 \sum_{i<j<k<l} V_{ij}
(\mathbf{r}_{ij})
  V_{kl}(\mathbf{r}_{kl}) + \ldots \, ,
\]
where $V_{ij}(\mathbf{r}_{ij})=\exp(-\beta \varepsilon_{ij} 
v_{ij}(\mathbf{r}_{ij}))$ is the Boltzmann factor associated with the
energy $\varepsilon_{ij}$ of making a bond between the $i$th and the $j$th
base at distance $\mathbf{r}_{ij}$. In this expression, $\beta= 1/T$ 
denotes the inverse temperature, and $v_{ij}(\mathbf{r}_{ij})$
represents the (short range) space dependent part of the interaction.
To further simplify the model, we will assume that
the chain is infinitely flexible and we will neglect all sterical
constraints, so that any pairing of bases is assumed to be feasible.
Therefore, we can neglect all spatial degrees of freedom and write
\begin{equation}
\mathcal{Z} = Z_{L}
=  1+  \sum_{i<j} V_{ij}
+ \sum_{i<j<k<l} V_{ij}
  V_{kl}+ \ldots \, ,
\label{ZL} 
\end{equation}
where now $V_{ij}=\exp(-\beta \varepsilon_{ij})$.
As shown in \cite{oz}, each term in
$Z_{L}$ can be represented graphically by a suitable arc diagram.  In
such a representation the nucleotides are dots on an oriented
horizontal line (which represents the RNA sugar backbone from the $5'$
end to the $3'$ end), and each base pair is drawn as an arc - above
that line - between the two interacting bases.  In real RNA, not all
pairs of nucleotides can interact. For instance, two bases which are
too close to each other along the backbone (say within a distance of
$4$ bases) cannot form a hydrogen bond since the backbone is not
flexible enough. Moreover, for an RNA molecule one also usually assumes
that only standard Watson-Crick pairs (A-U,C-G) and wobble pairs (G-U)
are possible. These
constraints greatly increase the difficulty of enumerating all
possible structures that are allowed. Among the set of all possible 
structures, one defines 
{\it secondary structures} of an RNA molecule as all structures which 
are represented by 
planar arc diagrams (no crossing of arcs). When the diagrams are non 
planar, one says that the RNA molecule contains one or more {\it 
pseudoknots}. Structures with pseudoknots can be classified 
according to the topological character of the corresponding arc 
diagram \cite{VOZ}.  Such a classification can be made more explicit 
directly in eq.~(\ref{ZL}), as explained in \cite{oz}. The main idea 
of \cite{oz} is to consider the following integral over matrices: 
\nopagebreak[4]\begin{eqnarray} 
Z_{L}(N)&=&\frac{1}{A_{L}(N)}\int \prod_{k=1}^{L}d\varphi _{k}\, e^{-\frac{N}{2}{ 
\sum_{ij}}(V^{-1})_{ij}\,{\rm tr}(\varphi _{i}\varphi _{j})} 
\nonumber \\ 
&& \times \frac{1}{N}  
{\rm tr}\prod_{l=1}^{L}(1+\varphi _{l})  \, . 
\label{matrep} 
\end{eqnarray} 
Here $\varphi _{i}$, $i=1,\ldots,L$, are $L$ independent $N \times N$ 
Hermitian matrices ($\varphi_{i}^{+}=\varphi_{i}$) and 
$\prod_{l=1}^{L}(1+\varphi_{l})$ is the ordered matrix product 
$(1+\varphi _{1})(1+\varphi _{2})\cdots (1+\varphi _{L})$. The 
normalization factor is:
\begin{equation} 
A_{L}(N)=\int \prod_{k=1}^{L}d\varphi _{k}e^{-
\frac{N}{2}{\sum_{ij}}(V^{-1})_{ij} 
{\rm tr}(\varphi _{i}\varphi _{j})} \, , 
\label{ALN} 
\end{equation} 
and $V$ is the $L\times L$ symmetric matrix with elements $V_{ij}$. 
The integral in eq.~(\ref{matrep}) can be evaluated
by using the Wick theorem.  The result is a function of $N$ which can 
be written as an asymptotic series at large $N$: 
\begin{eqnarray} 
Z_{L}(N) 
   &=&  1+  \sum_{i<j} V_{ij} + \sum_{i<j<k<l} V_{ij} V_{kl} \nonumber \\ 
&+& \frac{1}{N^{2}} \sum_{i<j<k<l} V_{ik} V_{jl}+ \ldots  \, . 
\label{ZLN} 
\end{eqnarray} 
The relation with the expansion in eq.~(\ref{ZL}) is obvious. The two 
series coincide for $N=1$, whereas for $N > 1$ the series in  
eq.~(\ref{ZLN}) contains topological information. All the planar structures 
are given by the $O(1)$ term of eq.~(\ref{ZLN}) and higher-order terms in 
$1/N^2$ correspond to RNA secondary structures with pseudoknots. The
classification of pseudoknots induced by this expansion is 
reviewed in \cite{VOZ}. 

The most challenging problem in RNA-folding prediction is to find
the structure with the lowest free energy.
If one restricts 
the search to the set of secondary structures without pseudoknots,
several fast algorithms are available \cite{mfold}. 
However, when one includes the
possibility of having pseudoknots, the problem is still
open.  An even simpler fundamental problem,
namely the exact combinatorics of RNA
structures with any pseudoknots, is unsolved. 
Results about the combinatorics of RNA secondary structures
without pseudoknots or with very special classes of pseudoknots
are available
(e.g.~\cite{combinatorics1,combinatorics2}), but the general case
is still lacking.  In this Letter we address precisely the problem
of enumerating all secondary structures with pseudoknots.
 
\begin{figure}[hbt]  
   \centering \includegraphics[width=0.4\textwidth]{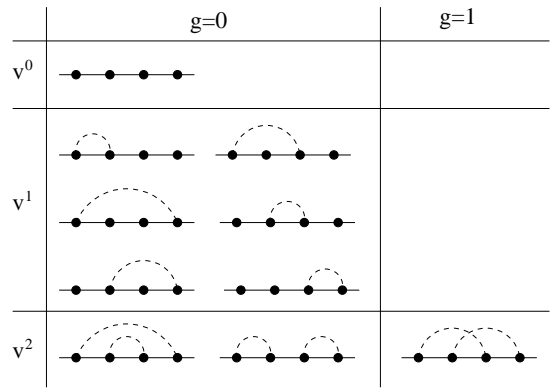}
   \caption{All possible arc diagrams with $L=4$. Diagrams with $i$
   arcs are associated to the power $v^i$, and $g$ is the genus.}
   \label{fig:L4}
\end{figure}

In order to get exact results, we make a few additional simplifications.
We assume that any possible pairing between nucleotides
is allowed (independently of the type of nucleotides and from their
distance along the chain) and that all the pairings may occur with
the same probability. In other words, we assume that the matrix
$V_{ij}$ has all entries equal $v>0$, i.e.:
\begin{equation} 
V=\left( 
\begin{array}{cccc} 
v+a&v&\cdots&v\\ 
v& v+a &\cdots&v \\ 
\cdots & \cdots &\cdots&\cdots \\ 
v& v& \cdots& v+a\\ 
\end{array} 
\right) \, . 
\label{V} 
\end{equation} 
The real number $a$ has been added in order for $V$ to be definite positive.
Of course this addition is purely formal since $Z_L(N)$ does not depend
on $a$, as one can easily see from eq.~(\ref{ZLN}). In fact no 
diagonal term $V_{ii}$ appears, as there are no 
self interaction diagrams. Even though the combinatorial problem in 
eq.~(\ref{ZL}) is now greatly simplified, it still
keeps a lot of its topological interest. In fact, by means of the matrix
integral in eq.~(\ref{matrep}) we can study the
distribution of RNA structures with pseudoknots as a function of their
topological character.  Let us illustrate this point by a simple
example for $L=4$. In this case all possible contact structures are 
listed in Figure \ref{fig:L4}.
 
There is a total of ten possible arc diagrams, nine of which are
planar and one which is not planar. The
nine planar diagrams contain one diagram without arcs, six with one
arc and two with two arcs. The same result can be directly obtained by
computing the matrix integral in eq.~(\ref{matrep}). In fact, as we will
show later in this Letter, the integral evaluates precisely to
$Z_{4}(N)=1+6v+2v^2+v^2/N^2$.  Thus the coefficients of the asymptotic
series have a direct topological interpretation, and that is the
reason why the asymptotic $1/N^2$ expansion is usually referred as {\it
topological expansion}
\cite{thooft}. Each term of the series gives the number of diagrams 
with fixed topological character: the first term represents planar 
diagrams, the second represents diagrams which can be drawn planarly on a 
surface with one handle (the torus), the third are diagrams that can be drawn 
planarly on a plane with two handles and so on. If we evaluate the 
integral in eq.~(\ref{matrep}) for any finite $L$ and finite $N$,
we will have an analytical control over the topology and the 
combinatorics of eq.~(\ref{ZL}). In the rest of the Letter, we will show 
explicitly how to compute the integral in eq.~(\ref{matrep}). 

First, we note that by using a series of Hubbard-Stratanovich
transformations, eq.~(\ref{matrep}) can be exactly simplified to:
\begin{equation} 
Z_{L}(N)=\frac{1}{\tilde{A}(N)} \int d\sigma \, 
e^{-\frac{N}{2v}\,{\rm tr} \sigma^2} 
\frac{1}{N} {\rm tr}(1+\sigma)^L  \, .
\label{gauss} 
\end{equation} 
We see that the original integration over the $L$ matrices $\varphi_k$ in
eq.~(\ref{matrep}) has been reduced to an integration over a single
$N \times N$ matrix $\sigma$. The similarity of eq.~(\ref{gauss}) and
(\ref{matrep}) is obvious, and will be demonstrated in a future publication (\cite{tobe}).
Note that the regulator $a$ drops out as long is it not zero. 
The  normalization factor $\tilde{A}(N)$ is: 
\begin{equation} 
\tilde{A}(N)= \int d\sigma \, 
e^{-\frac{N}{2v}\,{\rm tr} \sigma^2}= 
\left(
\frac{\pi v}{N}
\right)^{\frac{N^2}{2}}
2^{\frac{N}{2}}
 \, . 
\label{A1} 
\end{equation} 
The Gaussian matrix integral in eq.~(\ref{gauss}) is straightforward.
We introduce the spectral density of the matrix $\sigma$ at finite 
$N$: 
\begin{equation} 
\rho_N(\lambda) \equiv \frac{1}{\tilde{A}(N)} \int d\sigma \, 
e^{-\frac{N}{2v}\,{\rm tr} \sigma^2} 
\frac{1}{N} {\rm tr}\,  \delta(\lambda-\sigma) \, . 
\end{equation} 
By inserting the identity 
$1=\int_{-\infty}^{+\infty} d\lambda \, 
\rho_N(\lambda)$ into eq.~(\ref{gauss}), we obtain: 
\begin{equation} 
Z_{L}(N)= \int_{-\infty}^{+\infty} d\lambda \, \rho_N(\lambda) 
(1+\lambda)^L \,. 
\end{equation} 
Thus the multi-dimensional integral of eq.~(\ref{matrep}) has been reduced 
to a one-dimensional integral.  At this point it is convenient to 
study the exponential generating function of $Z_{L}(N)$: 
\begin{equation} 
G(t,N)\equiv \sum_{L=0}^{\infty} Z_{L}(N)\frac{t^L}{L!}= 
\int_{-\infty}^{+\infty} d\lambda \, \rho_N(\lambda) e^{t(1+\lambda)} \, . 
\label{G} 
\end{equation} 
The explicit form of $\rho_N(\lambda)$ is a well known and classic 
result of Random Matrix Theory (see e.g.~\cite{mehta} or 
\cite{BZ}, and we use it in the form given in \cite{ACMV}):
\begin{equation} 
\rho_N(\lambda)=\frac{e^{-\frac{N}{2v} \lambda^2}}{\sqrt{2 \pi v N}}  
\sum_{k=0}^{N-1} \left(  
\begin{array}{c} 
N\\ 
k+1 
\end{array} 
\right) 
\frac{H_{2k}(\lambda \sqrt{\frac{N}{2v}})}{2^k k!} \, , 
\label{rho} 
\end{equation} 
 where $H_k(x)$ are the Hermite polynomials: 
\begin{equation} 
H_k(x)=(-1)^k e^{x^2} \frac{d^k }{dx^k}e^{-x^2} \, . 
\end{equation} 
By inserting eq.~(\ref{rho}) into eq.~(\ref{G}), one obtains: 
\begin{equation} 
G(t,N)= 
\frac{1}{N}  \sum_{k=0}^{N-1}  
\left(  
\begin{array}{c} 
N\\ 
k+1 
\end{array} 
\right) 
\frac{(t^2 v)^{k}}{k! N^k} e^{\frac{v\,t^2}{2N}+t}\, , 
\label{almost}
\end{equation} 
where we have used the formula:  
\begin{equation} 
\int_{-\infty}^{+\infty} dx \, e^{-x^2+xy}H_n(x)= y^n  
e^{y^2} \sqrt{\pi}  \, . 
\end{equation} 
The sum in eq.~(\ref{almost}) can be expressed as a generalized Laguerre 
polynomial: \begin{equation} 
L^{(1)}_{N}(z)=\sum_{k=0}^{N}
\left(
\begin{array}{c}
N+1\\
N-k\\
\end{array}
\right) 
\frac{(-z)^k}{k!} \, .  
\label{hyper} 
\end{equation} 
We finally obtain: 
\begin{eqnarray} 
G(t,N)&=&e^{\frac{v \, t^2}{2N}+t} \frac{1}{N} L^{(1)}_{N-1}\left(-\frac{v 
\,t^2}{N} \right) \, . 
\label{G:hyper} 
\end{eqnarray} 
From this exact result we can extract informations on all the 
coefficients $Z_{L}(N)$. The series expansion 
in $t$ of $G(t,N)$ gives the first few coefficients $Z_L(N)$: 
\begin{center} 
\begin{tabular}{|c|l|} 
\hline 
$L$ & $Z_{L}(N)$ \\ 
\hline 
\hline 
1 & $1$ \\ 
\hline 
2 & $1+v$ \\ 
\hline 
3 &  $1+3v$\\ 
\hline 
4 & $1+6v+2v^2+v^2/N^2$\\ 
\hline 
5 & $1+10v+10v^2+5v^2/N^2$\\ 
\hline 
6 & $1+15 v+30v^2+5v^3+(15v^2+10v^3)/N^2 $\\ 
\hline 
7 & $1+21v+70 v^2+35 v^3+(35v^2+70v^3)/N^2 $\\ 
\hline 
8 & $1+28v+140v^2+140v^3+14 v^4+(70v^2+280v^3$ \\  
& $+70 v^4)/N^2+21v^4/N^4$\\ 
\hline 
\end{tabular} 
\end{center} 
The meaning of these values is straightforward: the power of $v$ is
the number of arcs in the diagram, and the power of $1/N^2$ is the
genus of the diagram. For instance when $L=7$ there are 21 planar
diagrams with one arc, and 35 diagrams on the torus (i.e.~genus one
closed oriented surface) with two arcs. The total number of diagrams
for each fixed genus can be obtained by putting $v=1$ (for instance,
the total number of diagrams on the torus for $L=6$ is
25). Analogously, the total number of diagrams, irrespective of
the genus, can be obtained by putting $N=1$ (for instance, the
number of diagrams for $L=4$ with 2 arcs is 3). 

\begin{figure*}[hbt]  
\includegraphics[width=0.4\textwidth]{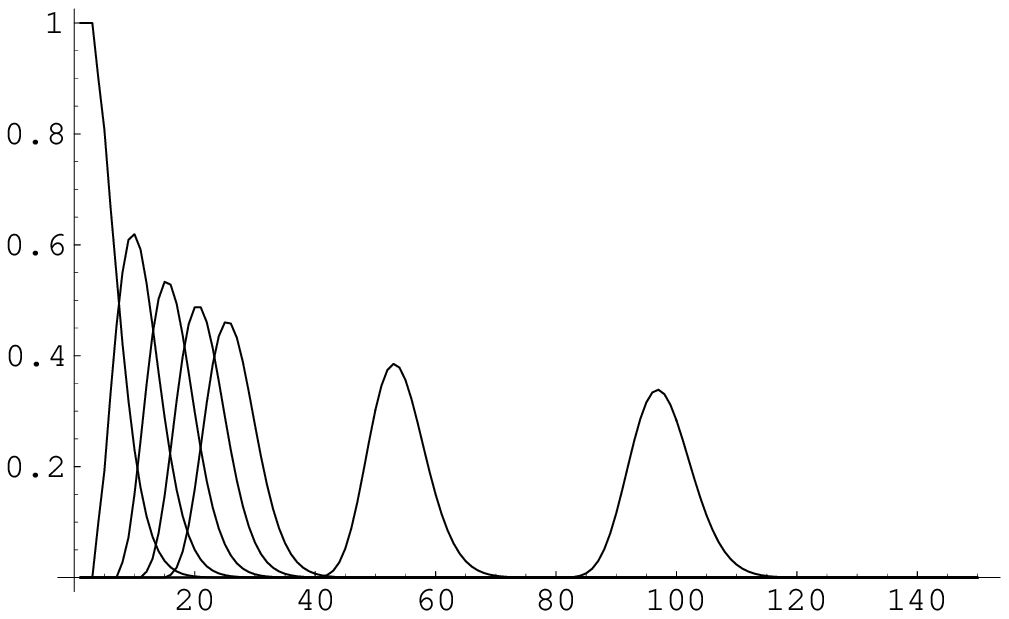} 
\put(-215,128){$a_{L,g}/{\cal N}$}
\put(-2,0){\small $L$}
\put(-180,122){\small $g=0$}
\put(-178,82){\small $g=1$}
\put(-170,72){\small $g=2$}
\put(-164,67){\small $g=3$}
\put(-153,60){\small $g=4$}
\put(-125,57){\small $g=10$}
\put(-70,50){\small $g=20$}
 \includegraphics[width=0.4\textwidth]{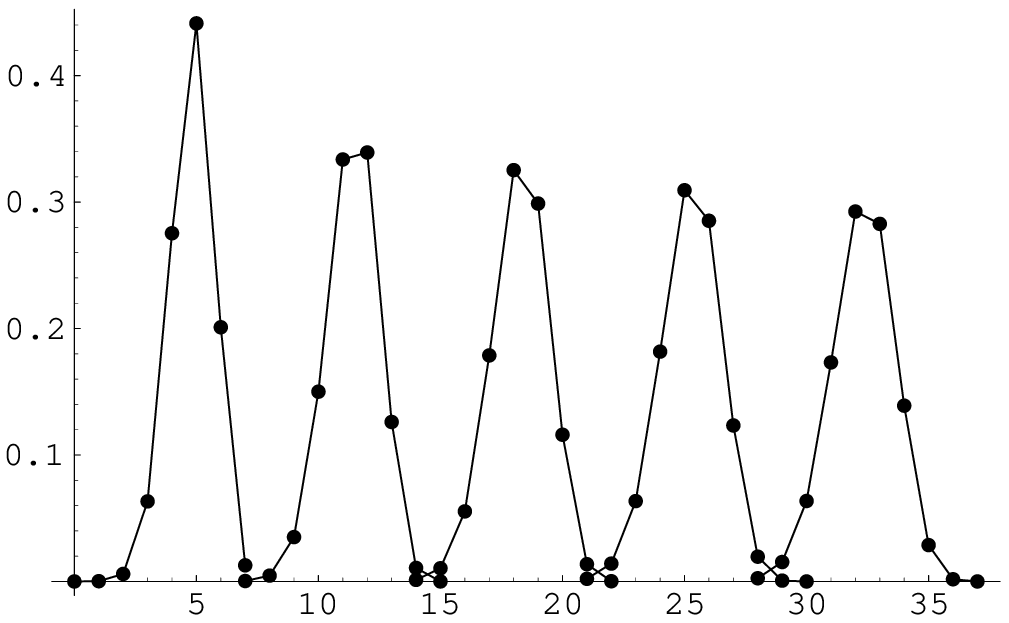} 
\put(-215,128){$a_{L,g}/{\cal N}$}
\put(0,1){\small $g$}
\put(-175,125){\small $L=30$}
\put(-140,99){\small $L=60$}
\put(-102,96){\small $L=90$}
\put(-64,91){\small $L=120$}
\put(-25,85){\small $L=150$}
\caption{On the left: the normalized number of diagrams $a_{L,g}/{\cal 
N}$ at fixed $g$  as a function of $L$.  On the right, the same 
quantity at fixed $L$ as a function of $g$. In both cases we put $ v=1$.}
\label{fig:aLg}
\end{figure*}

The general $1/N^2$ topological expansion of $Z_L(N)$ with $v=1$ is: 
\begin{equation}
Z_L(N)=\sum_{L=0}^{\infty} a_{L,g} \frac{1}{N^{2g}} \, ,
\label{topo}
\end{equation}
where the coefficients $a_{L,g}$ give exactly the number of diagrams
at fixed length $L$ and fixed genus $g$. From formula
(\ref{G:hyper}) and eq.~(\ref{topo}) we recursively obtain all the
coefficients $a_{L,g}$. Moreover, by normalizing each $a_{L,g}$ by the
total number of diagrams at fixed $L$, i.e.~by ${\cal N}\equiv
Z_L(1)$, we can obtain the distribution of the number of diagrams. In
Figure \ref{fig:aLg} we plot the distributions of diagrams as a
function of $L$ and $g$. We note the interesting feature that for any
given $L>>1$ most of the diagrams are not planar, and they
have a genus close to a characteristic value $\langle g
\rangle_L$. Such a value increases with $L$: we find numerically that it
scales like $\langle g \rangle_L \sim 0.23 L$, at large $L$.
Also, for each fixed $L$ there is a maximum possible
value for $g$, namely $g \leq L/4$. Conversely a structure can have a
genus $g$ only if it has a length at least $L\geq 4g$.

It is important to note that even though the number of planar
diagrams, $a_{L,0}$, is exactly the number of secondary structures
without pseudoknots, and the number of diagrams on a
torus, i.e.~$a_{L,1}$ counts structures with one pseudoknot only,
$a_{L,g}$ with $g\geq 2$ counts structures that contains either a single
topologically complex pseudoknot, or several simple pseudoknots 
with small genus. For that reason, the concept of {\it irreducible
pseudoknots} has been introduced in \cite{oz}, and it would be of interest to
study their distribution. The present analysis will be extended
to the case of irreducible pseudoknots in a future publication, where
we will also compute exact asymptotic behaviours for long sequences 
(\cite{tobe}).

In this Letter, we have shown how one can compute the 
number of folded structures as a function of the length and of the 
genus of the RNA. This model is of course very schematic and
oversimplified. It shows however that for a random RNA, 
the average topological
character scales linearly with the length of the chain. As most wild
RNA have an almost planar structure (with a genus $g \le 2$), this
implies that their sequences have been greatly designed by evolution
in order to achieve this specificity.

\underline{Acknowledgments}: 
We wish to thank G.~Cicuta, W.~Eaton, B.~Eynard, P.~Di Francesco,
E.~Guitter, S.~Nonnenmacher, L.~Molinari, for useful discussions. This
work was supported in part by the National Science Foundation under
grant number PHY 99-07949. GV acknowledges the support of the European
Fellowship MEIF-CT-2003-501547.

\end{document}